\documentclass[twocolumn,showpacs,preprintnumbers,amsmath]{revtex4}

\newcommand{\al}{\alpha}
\newcommand{\bt}{\beta}
\newcommand{\ka}{\kappa}
\newcommand{\ga}{\gamma}
\newcommand{\te}{\theta}
\newcommand{\be}{\begin{equation}}
\newcommand{\ee}{\end{equation}}
\newcommand{\n}{\label}
\newcommand{\no}{\noindent}
\begin{document}
\title{Form invariant transformations between n-- and
m-- dimensional flat Friedmann--Robertson--Walker cosmologies}
\author{Mauricio Cataldo}
\altaffiliation{mcataldo@ubiobio.cl}%0915
\affiliation{Departamento de F\'\i sica, Facultad de Ciencias,
Universidad del B\'\i o--B\'\i o, Avenida Collao 1202, Casilla
5-C, Concepci\'on, Chile.\\}
\author{Luis P. Chimento}
\altaffiliation{chimento@df.uba.ar} \affiliation{Departamento de
F\'\i sica, Facultad de Ciencias Exactas y Naturales, Universidad
de Buenos Aires, Ciudad Universitaria Pabell\'on I, 1428 Buenos
Aires, Argentina.}
\date{\today}
\begin{abstract}

We illustrate how the group of symmetry transformations, which
preserve the form of the n--dimensional flat
Friedmann--Robertson--Walker cosmologies satisfying Einstein
equations, acts in any dimension. This group relates the energy
density and the isotropic pressure of the cosmic fluid to the
expansion rate. The freedom associated with the dimension of the
space time yields assisted inflation even when the energy density
of the fluid is a dimensional invariant and enriches the set of
duality transformations leading to phantom cosmologies.

\end{abstract}
\maketitle %\preprint{APS/159-QED}

\section{Introduction}
Over the past several years there has been much interest in
examining cosmology in higher dimensions to see if the standard
four--dimensional Friedmann--Robertson--Walker (FRW) cosmology can
be recovered. The idea that our physical four--dimensional
Universe is embedded in a higher--dimensional spacetime has also
also attracted the attention of particle physicists and
astrophysicists. Theoretical motivation for such attempts can be
found within the framework of many theories of unification, among
them string, superstring and M theory, require extra spatial
dimensions to be consistent. Until today a number of important
solutions of Einstein equations in higher dimensions have been
obtained and studied, and they have led to important
generalizations and wider understanding of gravitational fields.
In this respect, of interest are the works on n-dimensional black
holes\cite{Myers}, Kaluza--Klein inflationary
cosmologies~\cite{Abbott}, circularly symmetric perfect
fluids~\cite{Garcia}, black holes on branes~\cite{Emparan}, and
recently, contributions on braneworld
scenarios~\cite{Lukas,Randall,Himemoto}.

It is interesting to note that some authors have also considered
phenomenological analysis in higher dimensional cosmology. For
example, the phenomenological analysis of five-dimensional
cosmology was stimulated by the work of Binetruy, Deffayet, and
Langlois~\cite{Binetruy}, and subsequently by the Randall-Sundrum
model~\cite{Randall}.

On the other hand, scalar fields play a crucial role in describing
cosmological models. In the standard big-bang theory such fields
are included for solving most of the problems found at very early
times in the evolution of the universe, and are called the
``inflaton" scalar field~\cite{Guth,Linde,Albrecht}. This scalar
field is characterized by its scalar potential.

At the same time, measurements of the luminosity--redshift
relations observed for the discovered type Ia supernovae with
redshift $z > 0.35$~\cite{Perlmutter,Garnavich}, indicate that at
present the universe is expanding with an accelerated fashion
suggesting a net negative pressure for the universe. One plausible
explanation of this astronomical observation is based on the
introduction of a scalar field, which is called the
``quintessence" or ``dark energy" scalar field.

Although these scalar fields are quite different in nature, there
are authors who think that the ``inflaton" and the ``quintessence"
fields might be of the same nature, in which a very specific
scalar potential form is used~\cite{Peebles}.

In Ref. \cite{lmath} it was shown that in several physical
problems the Einstein field equations for flat FRW cosmological
models and Bianchi I-type metric containing a scalar field can be
linearized and solved by writing them in invariant form. In all
these cases explicit use has been made of the non-local
transformation group. The symmetry transformations that preserve
the form of the Einstein equations introduce an alternative
concept of equivalence between different physical problems
\cite{Chimento1}. Cosmological models are equivalent when the
corresponding dynamical equations are form invariant under the
action of that group. Hence, it will be interesting to investigate
the consequences of this group when the dimension of the space
time is taken to be a free parameter of the theory. Notice that
the multidimensional point of view has been used in general
relativity to extract information or to endow with properties
fields and/or physical systems belonging to spaces of different
dimensions.

In this sense, the purpose of the present work is to illustrate
how a group of symmetry transformations acts on n-- and m--
dimensional flat FRW cosmologies which satisfy Einstein equations.
This group relates the energy density and the isotropic pressure
of the cosmic fluid {\it (source variables)} to the expansion rate
{\it (geometrical variable)} linking two different cosmologies,
one of which could be accelerated. Hence, even when the energy
density is a dimensional invariant we can get assisted inflation
\cite{Chimento1}-\cite{Chimento1'} driven by the freedom
associated with the dimension of the space time. In the case of
requiring the condition (10) of Ref.~\cite{Cataldo} the linked
cosmologies become identical, they share the same scale factor, or
there is a duality between contracting and superaccelerated
expanding scenarios associated with phantom cosmologies
\cite{Chimento2}-\cite{todos}, i.e. the scale factor of one of
them is the inverse of that of the other. The above formulation
also can be applied to a self--interacting scalar field using its
conventional perfect fluid description.

The outline of the present paper is as follows: In Sec. II we
review the well known Einstein equations for the FRW metrics in
n-- and m-- dimensional gravities coupled to a perfect fluid. The
case for constant barotropic indices is discussed in detail. In
Sec. III we briefly review the field equations for the FRW metrics
coupled to a scalar field. In Sec IV some conclusions are given.

\section{Dimensional form invariance symmetry in flat FRW spacetimes}
We shall assume the spherically symmetric flat FRW metric of an
$n$--dimensional spacetime given by
\begin{eqnarray}\label{ndim}
ds^2=-dt^2+a_{_{n}}(t)^2\left(dr^2+r^2 d\Omega_{n-2}^2\right),
\end{eqnarray}
where the spherical sector, related to $n-2$ angular variables
$\theta_i$, with $i$ running from $1$ to $(n-2)$, is determined to
be ${d \Omega_{_{n-2}}}^2={d \theta_{_{1}}}^2 + \rm{sin}^2
\theta_{_{1}}{d \theta_{_{2}}}^2 +...+\rm{sin}^2
\theta_{_{1}}...\rm{sin}^2 \theta_{_{n-3}} {d \theta_{_{n-2}}}^2$,
for $n \geq 3$. The Einstein equations for an $n$--dimensional
spacetime are given by
\begin{eqnarray*}
G_{_{\alpha \beta}}=R_{_{\alpha\beta}}-\frac{R}{2}
g_{_{\alpha\beta}}= \kappa_{_{n}} T_{_{\alpha\beta}},
\end{eqnarray*}
where Greek indices run from $1$ to $n$, and $\kappa_{_{n}}$
stands for the multidimensional gravitational constant.

The independent Einstein equations for the n--dimensional FRW
metric~(\ref{ndim}) filled with a perfect fluid are:
\begin{eqnarray}\label{1eq}
-{G_{t}}^{t}= \frac{(n-1)(n-2)}{2}
\frac{\dot{a}_{_{n}}^2}{a_{_{n}}^2} = \kappa_{_{n}} \rho_{_{n}},
\end{eqnarray}
\begin{eqnarray}\label{2eq}
-{G_{r}}^{r}=(n-2) \frac{\ddot{a}_{_{n}}}{a_{_{n}}}+
\frac{(n-2)(n-3)}{2}\frac{\dot{a}_{_{n}}^2}{a_{_{n}}^2} =
-\kappa_{_{n}} p_{_{n}},
\end{eqnarray}
where $\kappa_{_{n}}$, $a_{_{n}}$, $\rho_{_{n}}$ and $p_{_{n}}$
are the gravitational constant, scale factor, energy density and
the pressure in an n--dimensional spacetime respectively. Dots
denote differentiation with respect to $t$. The dependent Einstein
equations are related as ${G_{\theta_{_{n-2}}}}^{\theta_{_{n-2}}}
=...= {G_{\theta_{_{1}}}}^{\theta_{_{1}}}= {G_{r}}^{r}$.

We can replace Eq.~(\ref{2eq}) by the conservation equation:
\begin{eqnarray}\label{cemc}
\dot{\rho}_{_{n}}+(n-1) \, \frac{\dot{a}_{_{n}}}{a_{_{n}}} \,
(\rho_{_{n}}+p_{_{n}})=0,
\end{eqnarray}
which, as is well known, is derivable from the equation
${T^{\alpha \beta}}_{_{;\beta}}=0$. Thus the Einstein equations
for an n--dimensional flat FRW cosmology are given by
Eq~(\ref{1eq}) and Eq~(\ref{cemc}), which we shall rewrite in the
form:
\begin{eqnarray}\label{n1eq}
\alpha \, H^{\,2} = \kappa \, \rho, \qquad \dot{\rho}+\beta \, H
\, (\rho+p)=0,
\end{eqnarray}
where $\alpha=(n-1)(n-2)/2$, $H=\dot{a}/{a}$, $\beta=(n-1)$ and we
have omitted the subindex $n$.

For a different m--dimensional flat FRW cosmology the Einstein
equations are given by
\begin{eqnarray}\label{m1eq}
\bar{\alpha} \, \bar{H}^{\,2} = \bar{\kappa} \,
\bar{\rho}, \qquad \dot{\bar{\rho}}+\bar{\beta} \,
\bar{H}  \, (\bar{\rho}+\bar{p})=0,
\end{eqnarray}
where $\bar{\alpha}=(m-1)(m-2)/2$,
$\bar{H}=\dot{\bar{a}}/\bar{a}$,
$\bar{\beta}=(m-1)$,
  and $\bar{a}$, $\bar{\kappa}$,
$\bar{\rho}$ and $\bar{p}$ are the scale factor, gravitational
constant, energy density and the pressure in an m--dimensional
spacetime respectively.

By ``invariant form" we shall mean that the system of equations
(\ref{n1eq}) transform into Eqs.~(\ref{m1eq}) under the symmetry
transformations:
\begin{eqnarray}\label{GS1}
\bar{\rho}=\bar{\rho}(\rho),
\\
\label{tH}
\bar H=\pm \,  \theta  \sqrt{\frac{\bar\rho}{\rho}}\,H,
\\
\label{trp}
\bar\rho+\bar p=\pm\frac{\beta}{\bar\beta \, \theta} \sqrt{\frac{
\rho}{\bar\rho}}\, \frac{d\bar\rho}{d\rho}\,(\rho+p),
\end{eqnarray}
where $\theta=(\al\bar\ka/\bar\al \ka)^{1/2}$ and
$\bar{\rho}=\bar{\rho}(\rho)$ is an invertible function. Notice
that always $\te^2=\al\bar\ka/\bar\al \ka \geq 0$ since $n \geq 3$
and $m \geq 3$, and that these form invariant transformations are
defined without imposing any restriction on the cosmic fluid. When
the dimension of both cosmologies coincides, then we have
$\alpha=\bar{\alpha}$, $\beta=\bar{\beta}$, $\theta=1$, and these
transformations reduce to that of Ref.~\cite{Chimento1} and are
independent of the dimension where the cosmic fluid ``lives".

The invariant quantities associated with the set of
transformations~(\ref{GS1})-(\ref{trp}) are
\begin{eqnarray}\label{i1}
\frac{\bar\alpha\bar H^2}{\bar \kappa\bar\rho}= \frac{\alpha
H^2}{\kappa\rho},\\
\label{trpp}
\n{i2}
\frac{d\bar\rho}{\bar\beta\bar H(\bar\rho+\bar p)}=\frac{d\rho}{\beta H(\rho+p)}.
\end{eqnarray}

\no The first invariant expresses that the expansion of the
universe is proportional to the multidimensional gravitational
constant and to the energy density contained in the universe.
However, the expansion dims with the dimension of the space time
because it is proportional to the factor $1/\alpha$. The second
invariant expresses the fact that the transformations do not
modify the cosmic time.

In the case of considering perfect fluids with equations of state
$p=(\gamma-1) \rho$ and $\bar{p}=(\bar{\gamma}-1) \bar{\rho}$ in n
and m--dimensional spacetimes respectively, we conclude that the
barotropic indices $\gamma$ and $\bar{\gamma}$ transform as
\begin{eqnarray}\label{cond10}
\bar{\gamma}=\frac{\bar\rho+\bar
p}{\bar\rho}=\pm\frac{\beta}{\bar\beta \, \theta} \left(\frac{
\rho}{\bar\rho}\right)^{3/2} \, \frac{d\bar\rho}{d\rho}\, \gamma
\end{eqnarray}
under the symmetry transformations~(\ref{GS1})--(\ref{trp}). In
what follows, the upper and the lower signs will be referred to as
the $(+)$ and $(-)$ branches respectively.

These general form--invariant transformations relate cosmologies
in two different dimensions. For instance, they can be used for
generating a new m--dimensional FRW cosmology from a given
cosmology in (3+1)--dimensions (with $n=4$), or in
(2+1)--dimensions (with $n=3$), where a lot of them are known.

In this direction we investigate the consequences of a simple
example generated by the following transformation between
energy densities:
\begin{eqnarray}\label{16}
\bar\rho=b^2 \rho,
\end{eqnarray}

\no with $b$ a positive constant. Inserting the latter
in~(\ref{tH}) and~(\ref{cond10}) we find that $a$ and $\bar a$
are related to each other by
\begin{eqnarray}\label{177}
\bar a=a^{\pm b\theta},
\qquad b\theta=\pm\,\frac{\beta \gamma}{\bar\beta \bar\gamma},
\end{eqnarray}
where without loss of generality the constant of proportionality has been set equal to unity. Hence, the deceleration parameter $q(t)=-H^{-2}\ddot a/a$ transforms as
\be \n{q} \bar q=-1\pm \frac{1}{b\te}(q+1).
\ee

When the energy density is a dimensional invariant, i.e. for the
condition $\bar\rho=\rho$ or $b=1$, we get the relation
$\theta=\pm\beta \gamma/(\bar\beta \bar\gamma)$ which may be
interpreted as a constraint for the barotropic indices
$\bar\gamma$ and $\gamma$, since the pressures are not the same in
both dimensions~\cite{Nota}. In this case an expanding universe
with a positive deceleration parameter, (+) branch,  transforms
into an accelerated one if $\te$ is taken to be large enough. This
means that by adequately selecting the dimension of the space time
we can get assisted inflation. For instance, for constant $\ga$
and $\bar\ga$, the Einstein equations lead to power law solutions:
\begin{eqnarray*}
\bar a=t^{2/\bar\bt\bar\ga}=t^{\pm 2\te/\bt\ga}=a^{\pm\te},
\end{eqnarray*}
after using Eq.~(\ref{177}) for $b=1$. Then, for the (+) branch
and $2\te>\bt\ga$ we obtain an accelerated expansion.

It is interesting to investigate the choice $b\te=1$, because from
(\ref{177}) we have $\bar a=a^{\pm 1}$. Now we pay attention to
the $(-)$ branch since in this case the symmetry transformation
$a\to a^{-1}$ (duality) maps the initial singularity at $t=0$,
$a(0)=0$, into other kind of singularity $\bar a(0)=\infty$, i.e,
the scale factor $\bar a$ and the scalar curvature $\bar R$
diverge at a finite time. In particular, for $\ga>0$ the $(-)$
branch of the power law solution $\bar a=(-t)^{-2/\beta\gamma}$
defined for $t<0$ diverges in the future at $t=0$. This kind of
singularity dubbed ``big rip" is a characteristic of phantom or
ghost cosmologies. Hence, using the condition $\beta
\gamma=-\bar\beta \bar\gamma$, we obtain the relation between the
n-- and m-- dimensional flat FRW cosmologies:

\be \n{ph} \frac{\bar\ga\bar\ka}{m-2}=-\frac{\ga\ka}{n-2}, \ee
\no defining the phantom sector of our model.

The case of considering structural invariance of the scale factors
$a(t)$ in n-- and m--dimensional FRW cosmologies, i.e. dimensional
invariance of the scale factor as it was assumed in
Ref.~\cite{Cataldo}, corresponds to selecting the (+) branch of
the transformations~(\ref{GS1})--(\ref{trp}) generated by Eq.
(\ref{16}) with $b\te=1$. Thus, the energy densities corresponding
to n-- and m--dimensional FRW cosmologies are related by
$\bar\rho=\rho/\te^2$ and barotropic indexes transform as
\begin{eqnarray}\label{eqsct1}
(n-1)\, \gamma && \rightleftharpoons (m-1) \,\,\bar\ga,
\end{eqnarray}
where we have used the notation of Ref.~\cite{Cataldo}. Note that
the results we have obtained by applying the
transformations~(\ref{GS1})--(\ref{trp}) enlarge those of
Ref.~\cite{Cataldo} and add the duality between contracting and
expanding cosmologies through the $(-)$ branch of the
transformations which was not considered in the previous paper.

Finally, when the dimension of both cosmologies coincides we have
$\alpha=\overline{\alpha}$, $\beta=\overline{\beta}$, $\theta=1$,
and Eqs.~(\ref{GS1})--(\ref{trp}) (or Eqs. (6)--(8) of
Ref.~\cite{Chimento1}) are independent of the dimension where the
cosmic fluid ``lives".

%%%%%%%%%%%%%%%%%%%%%%%%%%%%%%%%%%%%%%%%%
\section{The scalar field case}
%%%%%%%%%%%%%%%%%%%%%%%%%%%%%%%%%%%%%%%%%

Let us consider a self--interacting scalar field $\phi$ driven by a potential $V(\phi)$ having an associated perfect fluid energy tensor with energy density and pressure given by

\begin{eqnarray} \n{pr}
\rho=\frac{1}{2} \, \dot{\phi}^2+V(\phi), \qquad p=\frac{1}{2} \,
\dot{\phi}^2-V(\phi),
\\
\label{rho en m}
\bar{\rho}=\frac{1}{2} \,
\dot{\bar{\phi}}^2+\bar{V}(\bar{\phi}), \qquad
\bar{p}=\frac{1}{2} \,
\dot{\bar{\phi}}^2-\bar{V}(\bar{\phi}),
\end{eqnarray}
in n--and m--dimensional FRW space respectively. Now, Eqs.
(\ref{pr})-(\ref{rho en m}) along with
Eqs.~(\ref{GS1})-(\ref{trp}) give the rules of transformation
for $\phi$ and $V$
\begin{eqnarray}
\dot{\bar{\phi}}^2= \bar{\rho}+\bar{p}=
\pm\frac{\beta}{\bar\beta \, \theta} \, \sqrt{\frac{
\rho}{\bar\rho}} \, \frac{d\bar\rho}{d\rho}\, \dot{\phi}^2,
\end{eqnarray}
\begin{eqnarray}
\bar{V}=\bar{\rho} \mp \frac{1}{2} \,
\frac{\beta}{\bar\beta \, \theta} \, \sqrt{\frac{ \rho}{\bar\rho}}
\, \frac{d\bar\rho}{d\rho}\, \dot{\phi}^2.
\end{eqnarray}
To illustrate an application of the latter we assume  the
transformation~(\ref{16})-(\ref{177}), then we
obtain
\begin{eqnarray}\n{tf}
\dot{\bar{\phi}}^2= \pm \, \frac{\beta b}{\bar\beta \, \theta}
\,  \dot{\phi}^2,
\end{eqnarray}
\begin{eqnarray}\n{tv}
\bar{V}
=\frac{b}{2}\,\left[b\mp\,\frac{\beta}{\bar\beta\,\theta}\right]
\dot{\phi}^2+b^2V(\phi).
\end{eqnarray}

\no In addition, the scale factor transforms according to Eq.
(\ref{177}). Notice that the $(+)$ branch gives the m--dimensional
analog of the n--dimensional original cosmological model, while
the $(-)$ branch leads us, as in the previous section, to phantom
cosmologies. The transformed scalar fields are related to the
original one by a dimensional generalization of the transformation
considered in Ref.~\cite{Chimento2}. It represents a
generalization of the Wick rotation.

There is an interesting case to be investigated, for instance, let
us consider the restricted group of transformations defined by the
condition $\bar V\propto V$. Then, from Eq.~(\ref{tv}) we get
$V\propto \dot\phi^2$ and $\rho\propto\dot\phi^2$. In this case
Eq.~(\ref{n1eq}) can be solved by assuming a power law scale
factor with a scalar field of the form $\phi\propto\ln {t}$. The
final
 solution is
\begin{eqnarray}
\n{a}
a=t^{2/\bt\ga},
\end{eqnarray}
\begin{eqnarray}
\n{v}
V=\frac{(2-\ga)\phi_0^2}{2\ga} \, e^{-2\phi/\phi_0},
\quad \phi=\phi_0\ln{t},
\end{eqnarray}

\no where $\phi_0=(2/\beta)(\alpha/\kappa\gamma)^{1/2}$. These
equations represent the dimensional generalization of the ordinary
exponential potential and its associated power law solutions. The
respective solution in the m-dimensional flat FRW space time is
obtained by inserting the above n-dimensional solution~(\ref{v})
into the transformations~(\ref{tf}) and~(\ref{tv}); so a
straightforward calculation gives \be \n{bv} \bar
V=\frac{(2-\bar\ga)\bar{\phi_0}^2}{2\bar\ga}\,e^{-2\bar\phi/\bar\phi_0},
\qquad \phi=\bar\phi_0\ln{t}, \ee where we have used
Eq.~(\ref{177}) and $\bar\phi_0=b\phi_0(\bar\ga/\ga)^{1/2}$. The
scale factor is given by Eqs.~(\ref{177}) and~(\ref{a}): \be
\n{ba} \bar a=t^{2/\bar\bt\bar\ga}=t^{\pm 2b\te/\bt\ga}=a^{\pm
b\te}.
%\bar a=t^{\pm 2b\te/\bt\ga}.
\ee

\no This example shows that the form invariant transformations can
be used to generate new cosmological solutions in an m-dimensional
gravity from a seed one in an n-dimensional gravity.

%%%%%%%%%%%%%%%%%%%%%%%%%%%%%%%%%%%%%%%%%%%%%%%%%
\section{Conclusions}
%%%%%%%%%%%%%%%%%%%%%%%%%%%%%%%%%%%%%%%%%%%%%%%%%

The main goal of the present work is to illustrate how a group of
symmetry transformations acts on n-- and m-- dimensional flat FRW
cosmologies which satisfy Einstein equations. Cosmological models
are equivalent since the corresponding dynamical equations become
form invariant under the action of this group. For two different
cosmologies, i.e. n-- and m-- dimensional flat FRW metrics, this
group relates their energy densities, isotropic pressures and the
scale factors to generic dimensional parameters $\alpha$, $\beta$,
$\bar\alpha$ and $\bar\beta$. If the dimension of both cosmologies
coincides, i.e. $n=m$, then the group of symmetry transformations
relates their energy densities, isotropic pressures and the scale
factors only. In addition, a form invariant symmetry
transformation which violates the dominant energy condition
induces a duality between contracting and superaccelerated
expanding scenarios generating phantom cosmologies. All these
multidimensional considerations can also be formulated for the
scalar field associating a perfect fluid description with the
stress energy tensor.

Finally, these general form--invariance transformations
can be considered as an algorithm for generating a new
m--dimensional FRW cosmology from a known n--dimensional
cosmology. For instance, we can use as a seed solution one given
known cosmology in (3+1)--dimensional gravity, where there exist a
lot of solutions.

\section{acknowledgements}
One of the authors (MC) thanks A.A. Garc\'\i a for valuable
comments. The authors thank Paul Minning for carefully reading
this manuscript. This work was partially supported by CONICYT
through grants FONDECYT N$^0$ 1051086, 1030469 and 1040624 (MC),
and Project X224 by the University of Buenos Aires (LPC). It was
also supported by the Direcci\'on de Investigaci\'on de la
Universidad del B\'\i o--B\'\i o (MC) and Consejo Nacional de
Investigaciones Cient\'\i ficas y T\'ecnicas (LPC).


\begin{thebibliography}{xxxxx}
\bibitem{Nota} If we want to make them dimensionally
invariant, we must require that the barotropic indices be equal,
i.e. $\bar\gamma=\gamma$. Thus, the scale factors are related to
each other by $\bar a=a^{^{\pm \beta/\bar\beta}}$.
\end{thebibliography}
\end{document}